%% file: hdcg.tex
\documentclass[12pt,prd, aps, showpacs, groupedaddress, superscriptaddress,floatfix]{revtex4}

\input{pab}

\begin{document}
\bibliographystyle{apsrev}

\title{Hierarchically deflated conjugate gradient}

\author{P.~A.~Boyle}\affiliation{SUPA, School of Physics and Astronomy, The University of Edinburgh, Edinburgh EH9 3JZ, UK}

\collaboration{RBC and UKQCD Collaborations}

\noaffiliation{Edinburgh-2014/03}

\pacs{11.15.Ha, 
      11.30.Rd, 
      12.15.Ff, 
      12.38.Gc  
      12.39.Fe  
}

\date{\today}
\maketitle

\centerline{Abstract}

We present a multi-level algorithm for the solution of five dimensional chiral fermion formulations,
including domain wall and Mobius Fermions. The algorithm operates on the red-black preconditioned Hermitian operator, and directly 
accelerates conjugate gradients on the normal equations. The coarse grid representation of this matrix
is next-to-next-to-next-to-nearest neighbour and multiple algorithmic advances are introduced, which help
minimise the overhead of the coarse grid. The treatment of the coarse grids is purely four dimensional, and the bulk
of the coarse grid operations are nearest neighbour. The intrinsic cost of most of the coarse grid operations is
therefore comparable to those for the Wilson case. We also document the implementation of this algorithm in the BAGEL/Bfm
software package and report on the measured performance gains the algorithm brings to simulations at the physical point on 
IBM BlueGene/Q hardware.


\newpage

\section{Introduction}

The index theorem \cite{Atiyah,Brown} guarantees the existence of low modes of a Dirac operator in topologically
non-trivial $SU(N)$ gauge configurations, protected from zero only by the light quark mass. 
This poses a problem for Krylov solution
of the covariantly coupled Dirac equation since the condition number of the matrices involved
must necessarily diverge as $\frac{1}{a m_{ud}}$. This problem must arise for any lattice action
that possesses the correct continuum limit, since the density of these modes must be universal near the continuum limit.

Krylov sparse matrix inversions of these Dirac operators approximate the inverse of the Dirac
operator applied on the initial residual with a polynomial of the Dirac matrix. The coefficients
are chosen to minimise the residual under some norm, the details depending on the Krylov algorithm.
Worst case bounds on convergence rates for conjugate gradients (CG) \cite{Hestenes} can be obtained
using the Chebyshev minimax polynomials \cite{Saad}; the minimax error is dependent on the range over which
the Chebyshev approximation must be accurate, and as a result the convergence rate is determined by the condition
number
$\kappa = \frac{ \lambda_{\rm max}}{ \lambda_{\rm min}}$ 
of the Dirac matrix, with the ratio $\sigma$ of successive residuals given by

\beq
\sigma = \frac{\sqrt{\kappa} - 1}{\sqrt{\kappa} +1}.
\label{eq:convergence}
\eeq

This convergence factor becomes unity in the limit of large condition number, and convergence necessarily becomes slow.
However, if we suppose there is a physical density of instantons and anti-instantons, 
and that we are near the continuum limit, one expects a set of detached set of low virtuality eigenvectors of 
topological origin that could be removed from the problem via deflation \cite{Morgan:2001qr,Darnell:2003mi,Darnell:2007dr,Stathopoulos:2007zi}.
After deflation, the convergence rate is determined by the improved condition number corresponding to the rest of the
spectrum. As the number of these topological modes is proportional to the volume, the cost of the deflation is $O(V^2)$. 
Scaling to the large volume limit therefore imposes considerable cost. 

Problematic volume scaling has, in other areas of computational science, 
been alleviated to $O( V \log V)$ using multi-level algorithms \cite{Brandt}.
Gauge freedom has until recently posed a barrier in Lattice theory: the process of blocking gauge dependent variables
must be gauge covariant and hence discovered on each configuration, 
and it is only recently that two broadly similar approaches have been successfully developed.

Luscher\cite{Luscher:2007se} introduced the key concepts 
of local coherence of low modes; the low modes bear substantial local
similarity. He defined blocked variables in a gauge covariant way, using inverse iteration of the lattice Dirac operator to produce
vectors with near vanishing covariant derivative so that they are rich in low modes. Similar ideas
were independently introduced in the smoothed aggregation algebraic multigrid context\cite{Brezina}. 
The gauge symmetry in QCD is likely a worst case example of the case of non-smooth matrix coefficients considered in the algebraic
multigrid case. Luscher's idea of local coherence is similar to the ``weak approximation'' property,
and a vector posessing near vanishing covariant derivative is termed ``algebraically smooth'', with blocked variables being
named aggregates. This multigrid work propagated directly into the programme carried out in the US by
a group spanning both particle physicists and the mathematical community \cite{Brannick:2007ue,Brannick:2007cc,Clark:2008nh,Babich:2009pc,Babich:2010qb}

Most successful applications of this class of techniques to lattice QCD have been based on solution of the non-Hermitian
system for Wilson and clover Fermions \cite{Luscher:2007se,Brannick:2007cc,Babich:2010qb,Osborn:2010mb,Frommer:2012mv,Frommer:2013fsa,Frommer:2013kla}.

The methods differ in whether and how the blocked representation of the Dirac
operator is introduce as a preconditioner, whether more than two levels are included,
and in the method used to discover the relevant subspace. Some earlier effort has been invested in unpreconditioned domain
wall Fermions using the normal equations\cite{Cohen:2012sh}.

The purpose of this paper is to develop a similar method applicable to domain wall Fermions and other forms
of five dimensional Cayley form chiral Fermion 
\cite{Kaplan:1992bt,Shamir:1993zy,Furman:1994ky,Narayanan:1993sk,Narayanan:1993ss,Narayanan:1994gw,Neuberger:1997bg,Neuberger:1997fp,
Borici:1999da,Borici:1999zw,Kikukawa:1999sy,Brower:2004xi,
Edwards:2005an,Brower:2012vk}.
This five dimensional  approach is somewhat more challenging because 
the negative bulk Wilson mass shifts the spectrum of the 5d Wilson operator such that there are modes
in all four quadrants of the complex plane. Spectra not contained within an open half plane are
known to lead to difficulties for non-Hermitian Krylov methods\cite{trefethen}
since we must make a polynomial approximation to $\frac{1}{z}$ that must cover a region of the complex plane that contains
the pole. Domain wall Fermions indeed follow this folklore.
In terms of practical algorithms, to the author's knowledge, almost
all algorithms for 5d Cayley chiral Fermions have made use of the normal equations with one exception.
\footnote{The interesting exception is that the unsquared system has been rendered tractable on QCD configurations by transforming
to a real-indefinite spectrum by preconditioning with $\gamma_5 R_5 $,  using MCR-$\gamma_5 R_5$ introduced by 
Arthur \cite{Arthur}. Although convergent, there was no significant algorithmic speedup. This outer solver may, however,
be worth further investigation as the basis of a multi-level algorithm since the coarse operator will have a smaller stencil.}

Since one of the most effective Krylov solvers for DWF is CG\footnote{Arthur\cite{Arthur} also showed that MCR achieves similar efficiency} 
applied to the normal equations for the red-black preconditioned operator, we aim to start from this most efficient
point and in this paper develop and even faster algorithm based on inexact deflation. Since this operator
is next-to-next-to-next-to-nearest neigbour the coarsened operator is considerably 
more non-local than for the Wilson operator and multiple algorithmic advances were required to obtain a 
real, measured speedup.

In section \ref{sec:review} we review the background of blocked deflation spaces that underlies both
inexact deflation and adaptive multigrid.
The algorithm we introduce will be hierarchical, however initially in section~\ref{sec:twolevel}
we shall consider only the two level approach
and where one can initially assume for the sake of argument that all inner, coarse matrix inversions are performed exactly.
Once the two level approach is established we will then consider in Section~\ref{sec:ldop}
introducing a second level when the application of the deflation matrix inverse,
is performed in an inexact numerical implementation.
We will study the performance of the algorithm in section~\ref{sec:results}.

\section{Blocked deflation subspace}
\label{sec:review}
In this section we introduce our notation for the building blocks of multi-level algorithms.
To avoid critical slowing down in sparse matrix inversion, one can treat a vector subspace $S = \rm{span} \{ \phi_k \}$ exactly.
If the lowest lying eigenmodes are completely contained in $S$ the ``rest'' of the problem avoids critical slowing down.
The method requires some setup since in a gauge theory this subspace is gauge dependent:
one first must generate subspace vectors $\phi_k$ that are ``rich'' in low modes.
We address in section \ref{Sec:subspace} some procedures for how these vectors might be generated.
Luscher observed that
by subdividing these vectors into blocks $b$

\beq
\phi^b_k(x) = \left\{ \begin{array}{ccc} 
  \phi_k(x) &;& x\in b\\
  0 &;& x \not\in b
\end{array}
\right.
\eeq 
we obtain a  much larger subspace that, due to local coherence, is enormously more effective for deflation. 
These blocked vectors are locally orthogonalised 
using a Gramm Schmidt procedure. For example, on a 
$48^3\times 96$ lattice with $4^4$ blocks there is a 
$12^3\times 24$ coarse grid, and we obtain an $O(10^4)$ bigger deflation space than one
would expect from using the vectors $\phi_k$ without blocking. If the low modes are locally coherent,
the span of block subvectors will better contain the low mode subspace than that of the global vectors
because
\beq{\rm span} \{ \phi_k\}\subset 
{\rm span} \{ \phi_k^b\} .\eeq
We introduce blocked subspace projectors
\beq
P_S =  \sum_{k,b} |\phi^b_k\rangle \langle \phi^b_k | \quad\quad ; \quad\quad P_{\bar{S}} = 1 - P_S 
\eeq
and compute $M_{ss}$ as
\beq
M=
\left(
\begin{array}{cc}
M_{\bar{S}\bar{S}} & M_{S\bar{S}}\\
M_{\bar{S}S} &M_{SS}
\end{array}
\right)=
\left(
\begin{array}{cc}
P_{\bar{S}} M P_{\bar{S}}  &  P_S M P_{\bar{S}}\\
 P_{\bar{S}} M P_S &   P_S M P_S
\end{array}
\right)
\eeq
We can represent the matrix $M$ exactly on this subspace by computing its matrix elements, 
known as the \emph{little Dirac operator} (coarse grid matrix in multi-grid)
\beq
A^{ab}_{jk} = \langle \phi^a_j| M | \phi^b_k\rangle
\quad\quad ; \quad\quad
(M_{SS}) = A_{ij}^{ab} |\phi_i^a\rangle \langle \phi_j^b |.
\eeq
the subspace inverse can be solved by Krylov methods and is:
\beq
Q =
\left( \begin{array}{cc}
0 & 0 \\ 0 & M_{SS}^{-1}
\end{array} \right) 
\quad\quad ; \quad\quad
M_{SS}^{-1} = (A^{-1})^{ab}_{ij} |\phi^a_i\rangle \langle \phi^b_j |
\eeq
It is important to note that $A$ inherits a sparse structure from $M$ because well separated blocks do \emph{not} connect through $M$.
We can Schur decompose the matrix
\begin{eqnarray*}
M= U D L = \left[ \begin{array}{cc}M_{\bar{s}\bar{s}} & M_{\bar{s}s} \\ M_{s\bar{s}} & M_{ss} \end{array} \right]
&=&
\left[ \begin{array}{cc} 1 & M_{\bar{s} s}  M_{ss}^{-1} \\ 0 & 1 \end{array} \right]
\left[ \begin{array}{cc} S & 0 \\ 0 & M_{ss} \end{array} \right]
\left[ \begin{array}{cc} 1 & 0 \\ M_{ss}^{-1} M_{s \bar{s}} & 1 \end{array} \right]
\end{eqnarray*}
Note that 
$P_L M = \left[ \begin{array}{cc} S & 0 \\ 0 &0 \end{array} \right]$ yields the Schur complement $ S = M_{\bar{s}\bar{s}} - M_{\bar{s}s} M^{-1}_{ss} M_{s\bar{s}} $,
and that the diagonalisation $L$ and $U$ are related to Luscher's projectors $P_L$ and $P_R$ (Galerkin oblique projectors in multi-grid)
\beq
P_L = P_{\bar S} U^{-1} =\left( \begin{array}{cc}
 1 & -M_{\bar{S} S}  M_{SS}^{-1}\\
 0 & 0 
             \end{array} \right) 
\quad\quad ; \quad\quad
P_R = L^{-1} P_{\bar{S}}  = 
\left( \begin{array}{cc}
1 & 0 \\ -M_{SS}^{-1} M_{S \bar{S}} & 0  
\end{array} \right) 
\eeq
Finally, we require the relation $Q M = 1- P_R$. With a Hermitian system we gain the properties
\beq
P_L^\dagger = P_R 
\quad\quad\quad
(P_L M)^\dagger = P_L M
\eeq

\subsection{Schur Complement Algorithms}

Luscher introduced a class of algorithms based on the Schur decomposition. If we multiply the equation
\beq M\psi = \eta \eeq by $1-P_L$ and $P_L$ to obtain two equations yielding $(1-P_R)\psi$ and $P_R \psi$, we have:
\begin{eqnarray}
(1-P_R) \psi &=& M_{ss}^{-1} \eta_s \label{eq:easy}\\
(P_L M) \chi &=&  P_L \eta \label{eq:hard}\\ 
\psi &=& P_R \chi + M_{ss}^{-1} \eta_s 
\end{eqnarray}
Solving Eq.~\ref{eq:easy} is easy, while for Eq~\ref{eq:hard} each step of an outer Krylov solver involves an \emph{inner} Krylov solution of the little Dirac operator,
entering entering the matrix $P_L M$ being inverted. Any errors in this little Dirac operator
propagate into solution. Luscher alleviated this by tightening the precision during convergence, and using the history forgetting \emph{flexible} GCR
algorithm. The overhead of the little Dirac operator is suppressed by introducing the Schwarz alternating procedure (SAP) preconditioner as follows:
\beq
(P_L M)  M_{SAP} \phi = P_L \eta \quad\quad ;\quad \quad
\psi =  M_{SAP} \phi
\eeq
In Luscher's approach to Wilson fermions the little Dirac operator  for $D_W$ is \emph{nearest neighbour}, and although the fine
operator does not make use of red-black preconditioning,
red-black preconditioning of the little Dirac operator possible because the spatial structure
is preserved in the coarse operator. This means that the Schwarz alternating procedure remains
possible as $D_W$ does not connect red to red.

\section{Adaptive multigrid methods}

Multigrid algorithms are typically expressed in a more heuristic manner than pure Krylov solvers.
The underlying basic block is an error reduction step, taken with a (cheap) approximate inverse $\tilde{M}$:
\beqa
r  &=& \eta - M \psi\\
\psi^\prime &=& \psi + \tilde{M} r 
\eeqa
\beqa
\Rightarrow r^\prime &=& (1 - M \tilde{M}) r
\eeqa
To the extent that $\tilde M$ is a good approximate inverse a convergent process
can be built from these steps. The adaptive multigrid 
\cite{Brannick:2007ue,Brannick:2007cc,Clark:2008nh,Babich:2009pc,Babich:2010qb,Osborn:2010mb,Frommer:2012mv,Frommer:2013fsa,Frommer:2013kla}
for QCD is well developed for Wilson and clover Fermions. They typically combine an inner V-cycle or W-cycle 
spanning representations of the Dirac matrix on multiple grids as a preconditioner to an outer Krylov solver, and since non-Hermitian
systems are treated this is typically flexible GCR solver. As with Luscher's algorithm they typically adopt the SAP procedure as
a preconditioner, but this is organised as an error reducing smoother entering as one component in a sequence of error reduction 
steps on a multigrid cycle. The inclusion of these steps as a variable preconditioner is important since this allows composition when each
each step is an aggressively truncated Krylov process bearing substantial variability.
For Hermitian (symmetric) matrices a symmetric V-cycle is required for conjugate gradients, consisting of both pre- and post-smoothing steps
to ensure Hermiticity. In later sections we will see how this can be avoided.

\section{Generalisation to 5d Chiral Fermions}
\label{sec:twolevel}
To generalise Luscher's approach, we must implement a method suitable for Krylov solution of the Hermitian system.
In this work we will aim to speed up solution for the red-black preconditioned system, as this starting point is
the best presently known approach and so any speed up over the baseline is a genuine gain. We define the 
Hermitian red-black operator ${\cal H}$ as:
\beq
{\cal H} = \left( M_{oo} - M_{oe} M_{ee}^{-1} M_{eo} \right)^\dagger\left( M_{oo} - M_{oe} M_{ee}^{-1} M_{eo} \right)
= M_{\rm prec}^\dagger M_{\rm prec}
\eeq
The Hermitian naure of this matrix is an advantage for subspace generation, which we discuss below.
There are also several significant challenges that arise because
the operator is is next-to-next-to-next-to-nearest-neighbour in four dimensions, and entirely non-local in the fifth dimension.
Firstly we address the non-locality as follows.
We do not block the fifth dimension to address the non-locality.
In four dimensions the matrix stencil still connects 320 neighbours compared to the eight for 
the Wilson non-Hermitian system; a substantial suppression of the 
little Dirac operator overhead must be found to alleviate this additional cost.
Secondly, since the matrix is not nearest neighbour the alternating procedure cannot be applied;
we require an alternative to the Schwarz preconditioner. Thirdly,
we must find an appropriate solver:  $(P_L M) M_{SAP}$ is a non-Hermitian matrix so is unsuitable for
Hermitian solver algorithms such as conjugate gradients.
Finally we must ensure the system is tolerant to loose convergance of the inner Krylov solver(s); this
is needed to relax the stopping conditions, similar to the flexibility introduced in Luscher's algorithm.
We address these issues in turn.

\subsection{Preconditioned conjugate gradient for Schur complement}

\begin{figure}[hbt]
\begin{center}
\fbox{
\begin{minipage}{0.4\textwidth}
{\small
\begin{enumerate}
\item $r_0 = b - A x_0$
\item $z_0 = M_{IRS} r_0$ ; $p_0=z_0$
\item for iteration $k$
\item \hspace{1cm}$\alpha_k = (r_k,z_k)/(p_k,A p_k)$
\item \hspace{1cm}$x_{k+1}= x_k + \alpha_k p_k$
\item \hspace{1cm}$r_{k+1}= r_k - \alpha_k A p_k$
\item \hspace{1cm}$\mathbf{z_{k+1} = M_{IRS} r_{k+1}}$
\item \hspace{1cm}$\mathbf{\beta_k = (r_{k+1},z_{k+1})/(r_k,z_k)}$
\item \hspace{1cm}$\mathbf{p_{k+1} = z_{k+1} + \beta_k p_k}$
\item end for
\end{enumerate}
}
\end{minipage}
}
\end{center}
\caption{\label{fig:pcg}{\bf Preconditioned conjugate gradient algorithm}. In the iteration
both $M_{IRS}$ and $A = P_L M$ contain inner Krylov solves. However $M_{IRS}$ only enters the selection
of the search direction $p_k$, while the little Dirac inversion directly enters the linear combination coefficient
$\alpha_k$. We observe a drastically different sensitivity to the precision of the two inner inversions.}
\end{figure}

We initially applied the standard preconditioned CG \cite{AxelssonPcg,p278Saad} given in
figure~\ref{fig:pcg}
to the Schur complement operator 
\beq 
P_L {\cal H} =
\left( \begin{array}{cc}
 1 & -M_{\bar{S} S}  M_{SS}^{-1}\\
 0 & 0 
\end{array} \right) {\cal H}.\eeq 
Each iteration used two inner Krylov solvers:
the little Dirac operator inversion \beq Q \equiv M_{SS}^{-1}\eeq entering $P_L$, and 
the IR shifted preconditioner inversion \beq M_{IRS}=\frac{1}{{\cal H}+\lambda}.\eeq
The convergence precision on both of these is independently controllable and we will shortly
study the sensitivity of the overall convergence to both these Krylov inversions independently.
Although this is not the final manifestation of the new algorithm, the initial simplicity
of preconditioned CG is helpful for framing the following discussion. This is particularly the case
since the structure is very similar to that of Luscher's algorithm\cite{Luscher:2007se}.

\subsection{Subspace generation}

\label{Sec:subspace}

Since we are dealing with a Hermitian positive definite operator, an efficient subspace
generation can be performed using a multi-shift solver to approximate a fourth order rational low-pass filter 
applied to Gaussian noise vectors $\eta_k$, without the need for inverse iteration.

\beq
\phi_k = R(\eta_k) \propto
\frac{1}{({\cal H}+\lambda)
({\cal H}+\lambda+\epsilon)
({\cal H}+\lambda+2\epsilon)
({\cal H}+\lambda+3\epsilon)}
\label{Eq:ratfilt}
\eeq

\begin{figure}[hbt]
\includegraphics[width=0.5\textwidth]{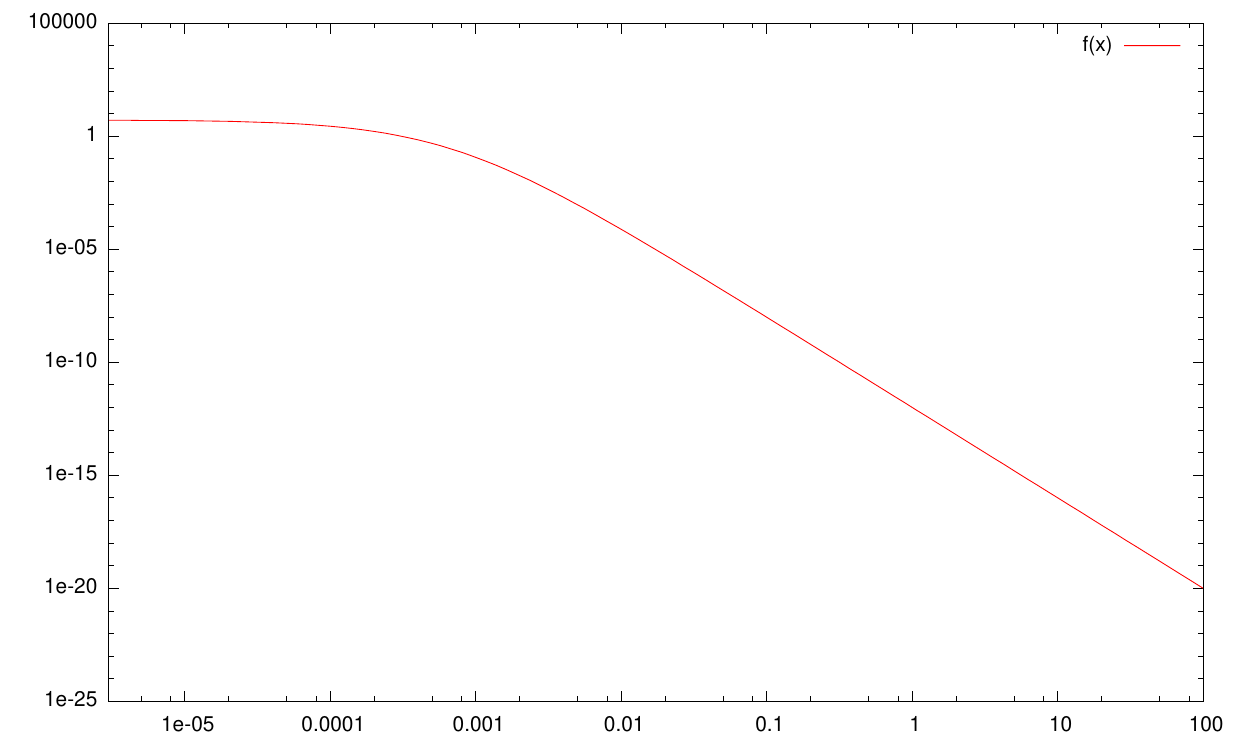}
\caption{\label{fig:bandpass} A single multishift inversion can well approximate a 4th order rational low pass filter
and be used in low mode subspace generation. The bandpass parameter $\lambda$ protects the condition number of the
inversions, while only a single pass is required. Together these make the process relatively cheap compared to 
inverse iteration. The filter function is given in Eq. \ref{Eq:ratfilt}, and we display the exact result obtained if the
multishift inversion was applied to very high convergence precision.
}
\end{figure}

\begin{figure}[hbt]
\includegraphics[width=0.5\textwidth]{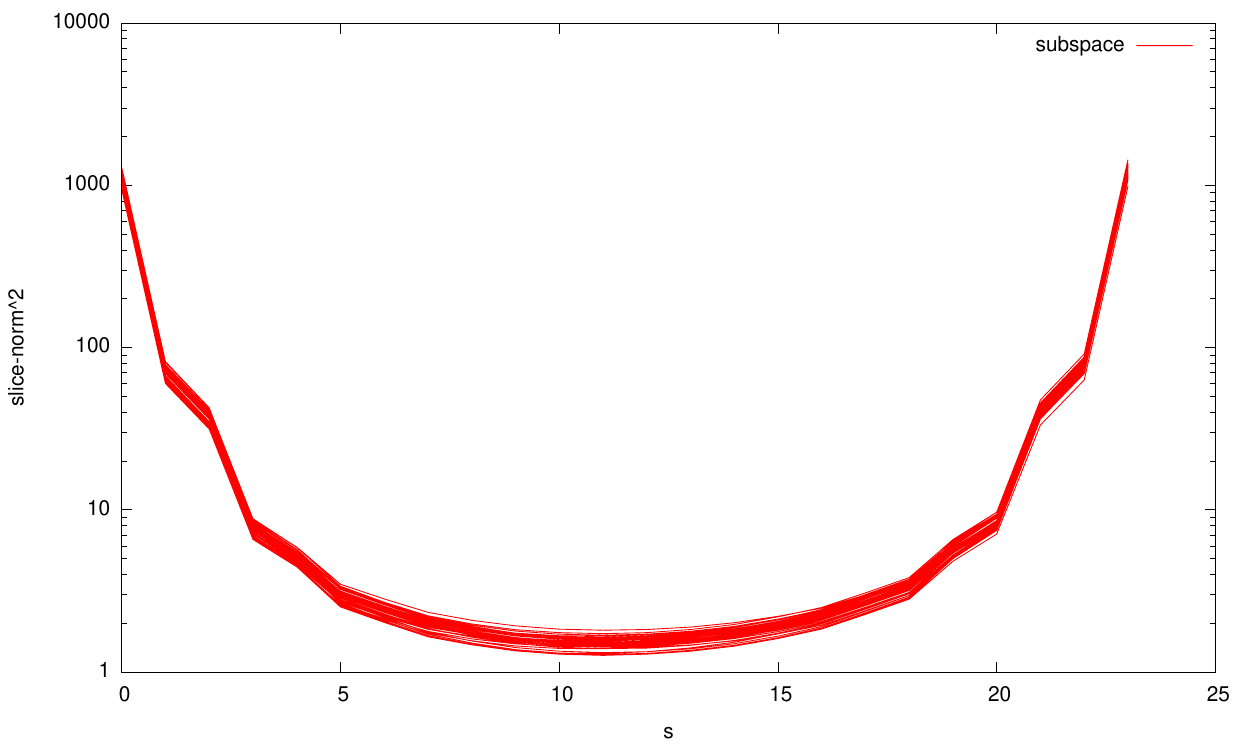}
\caption{\label{fig:subspace} Dependence of the norm of four dimensional slices of the subspace vectors
on the coordinate in the fifth dimension. Even for the preconditioned Hermitian Mobius Fermion case the
lowest modes are bound to the walls. All vectors obtained on a single configuration are overlayed, and the
shape profile is non-trivial. This shape profile can be used to our advantage in several ways (both in subspace generation
and accelerating the projection/promotion from the coarse spaces) since the interior
segments become negligible for large $L_s$.}
\end{figure}

Typically we take $\lambda = 0.0003$ and $\epsilon = \lambda/3$, however this depends on the normalisation 
and condition number of the operator. 
The response function of this rational filter is illustrated in figure~\ref{fig:bandpass}.
As one expects for domain wall Fermions the resulting vectors have non-trivial structure in the fifth dimension
being largely bound to the walls. This statement remains true for the preconditioned Hermitian Mobius Fermion case,
figure~\ref{fig:subspace}. 

This profile may be exploited in several ways.
A modest improvement in the quality of subspace was obtained by generating four dimensional 
Gaussian noise vectors $\eta_k$ and placing this only on the physical field components of the
walls as the input to our rational filter, thus better matching the input vector to these physical modes and
requiring less effort to be applied in the filtering. 

Two efficient strategies have been found, and which is more efficient depends on $L_s$. 
Firstly solving $R(\eta_k)$  to $10^{-6}$ is efficient for modest $L_s$. 
Secondly, we can generate the subspace vectors in two stages: a first pass approximation is generated 
by solving $R(\eta_k)$ with reduced precision and reduced $L_s$
ranging between eight and twelve.
These are promoted to the surface
regions of the full system, a second pass uses the first pass deflated solver to improve the subspace filling in the 
fifth dimension bulk. This second pass applies a single shifted inversion $ [ {\cal H} + \lambda ]^{-1} $ to the 
first pass subspace vectors (after promotion to the increased $L_s$) to some precision.

The code implementation can also optionally remove the interior
of a subset of the subspace vectors, recognising that they are near zero in this region. This limits
the growth of the cost of subspace generation, and of projection to and from the coarsened problem with
$L_s$, while the coarse problem cost does not depend on $L_s$.

BAGEL and the associate BFM (Bagel Fermion Matrix) package have an implementation of HDCG \cite{Boyle:2009bagel}.
The parameters in the BFM implementation of HDCG controlling subspace generation are tabulated in
Table~\ref{tab:FineSubspaceParams}.

\begin{table}[hbt]
\begin{tabular}{c|c}
Parameter & Meaning \\
\hline
\emph{SubspaceRationalLo} &  Low pass filter threshold $\lambda$ \\
\emph{SubspaceRationalLs} &  Extent of fifth dimension used in first pass subspace\\
\emph{SubspaceRationalResidual}  & Multishift solver convergence residual for first pass subspace\\
\emph{SubspaceRationalRefine} & Whether to generate a second pass subspace \\
\emph{SubspaceRationalRefineResidual} & Singleshift solver residual for second pass subspace\\
\emph{SubspaceRationalRefineLo} & Low pass filter threshold $\lambda$ for second pass subspace\\
\emph{SubspaceSurfaceDepth} & The depth in s-slices neighbouring the surface preserved in subspace vectors
\end{tabular}
\caption{\label{tab:FineSubspaceParams}
Parameters to HDCG for controlling the generation of the fine grid deflation space.
If a second pass is used the deflated solver generated by the first pass subspace is used to accelerate the 
generation of the second pass subspace. The second pass refinement consists of a single shift inversion with the
full five dimensional system.
}
\end{table}

\subsection{Infra-red shift preconditioner $M_{IRS}$ }

We aim to use outer iterations based on preconditioned CG (in a more modern form) 
\cite{AxelssonPcg,p278Saad}  with a Hermitian preconditioner.
A high mode preconditioner is required to
limit the cost overhead associated with the little Dirac operator.

In the multigrid literature this preconditioner would be called a \emph{smoother}, while
in Luscher's algorithm some fixed number of iterations of the Schwarz alternating procedure is used.
Since we are deflating the low modes, we seek approximate inverse preconditioner
for the Hermitian system that is both Hermitian positive definite and is a cost effective
approximate inverse reasonably accurate for high modes. 

BFM supports a number of options relating to the preconditioner for the outer Krylov process, listed in 
table~\ref{tab:preconditioner}.

\begin{table}[hbt]
\begin{tabular}{c|c|c}
Option & Value & meaning\\
\hline
\emph{PreconditionerType}&\emph{Mirs}      &  Fixed number of CG iterations with a infrared shift\\
&\emph{MirsPoly}  &  Fixed polynomial. Coefficients determined by one pass of CG\\
&\emph{Chebyshev} &  Chebyshev polynomial preconditioner\\
&\emph{None }     &  No fine grid preconditioner\\
\hline
\emph{PreconditionerKrylovIterMax} & int & Preconditioner iterations (or chebyshev order)\\
\emph{PreconditionerKrylovLo }     & double & Lowest eigenvalue targeted by preconditioner\\
\emph{PreconditionerKrylovHi }    & double & Upper limit of eigenrange for Chebyshev preconditioner
\end{tabular}
\caption{
\label{tab:preconditioner}
Options controlling outer Krylov preconditioner for HDCG implementation. \emph{PreconditionerType}
must be set to one of \emph{PreconditionerMirs}, \emph{PreconditionerMirsPoly},
\emph{PreconditionerChebyshev}, or \emph{PreconditionerNone}. 
\emph{PreconditionerKrylovIterMax} controls either the order of a polynomial or the number of Krylov
iterations as appropriate. \emph{PreconditionerKrylovLo} controls the eigenrange for both Krylov
based preconditioners and for the Chebyshev polynomial based preconditioner, while
\emph{PreconditionerKrylovHi } applies only to the Chebyshev preconditioner.
In practice we find that \emph{PreconditionerMirs} is the best option on physical point configurations.
}
\end{table}

The $M_{IRS}$ preconditioner is based on a fixed number of 
shifted CG iterations acting with a shifted matrix, and applied in single precision,
\beq
M_{IRS}({\cal H},N_{CG},\lambda) = \left. \frac{1}{{\cal H}+\lambda}\right|_{N_{CG} \rm iterations}.
\eeq
Here, $\lambda$ is a gauge covariant infra-red regulator that plays
a similar role to the domain size in SAP. Since a Krylov solver minimises the error of a
polynomial under some norm, the infra-red shift (IRS)
ensures the turning points of the error polynomial lie in the high eigenvalue region we wish 
to improve with the preconditioner. This ensures that the two preconditioners remain complementary.

\begin{figure}[hbt]
\includegraphics[width=0.5\textwidth]{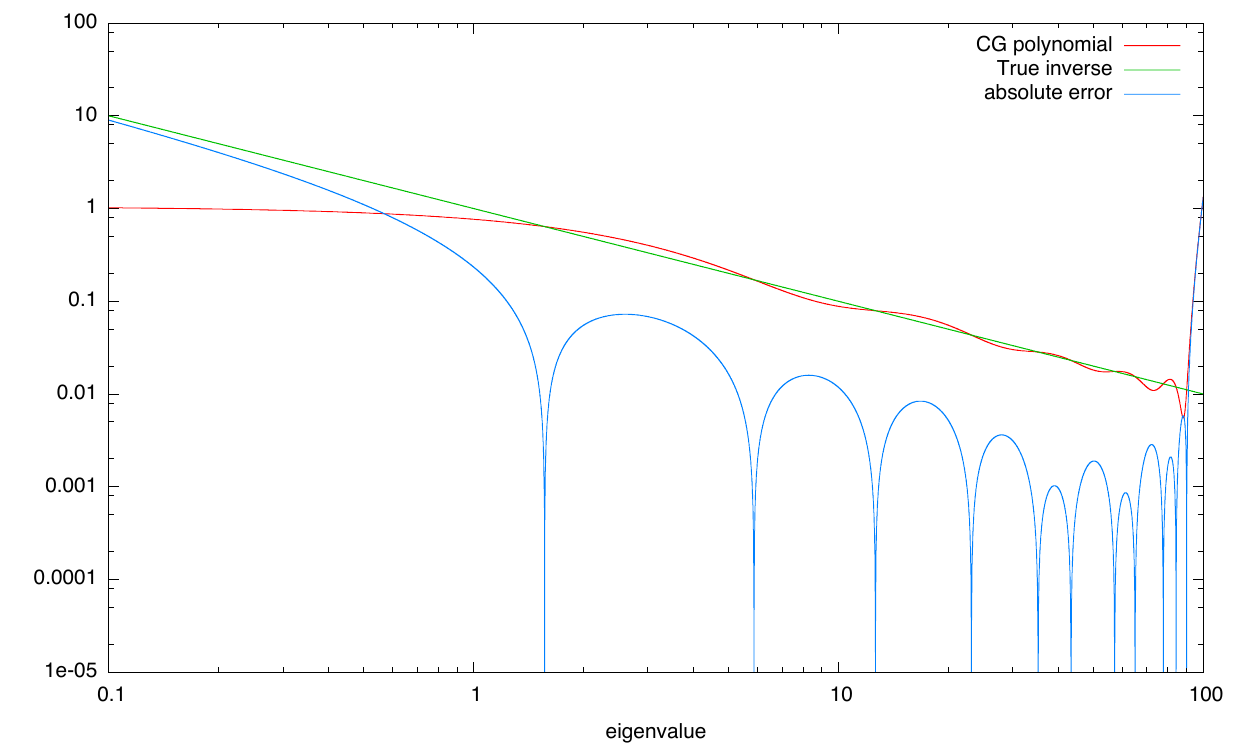}
\caption{
\label{fig:polycg}
Conjugate gradient approximate inverse polynomial selected by nine iterations of
conjugate gradient applied in $M_{IRS}$ with a low pass parameter $\lambda=1.0$.
Compared to the CG based polynomial preconditioner of \cite{Oleary} the inclusion of the infra-red
shift has a stabilising effect. The automatic selection of the upper edge of the spectrum (at around eigenvalue 
$\lambda\sim 90$) is an advantage
compared to Chebyshev preconditioners since the number of tunable parameters is reduced. In fact,
it turns out that the dynamic response of CG to the spectral content of the residual
is important; running a fresh Krylov process for each preconditioner application (\emph{PreconditionerMirs}) 
results in  faster convergence than freezing this with a static polynomial (\emph{PreconditionerMirsPoly}).
}
\end{figure}

A typical polynomial produced by the $M_{IRS}$ conjugate gradient may be seen in figure~\ref{fig:polycg}.
The lower range of the region of accuracy is set by the infra-red regulator $\lambda$. The upper
range is selected by the optimality of conjugate gradient, and thus detected by the algorithm
based on the spectrum of the operator and the spectral content of the initial residual.
In the example shown, the unpreconditioned operator has upper edge of its spectrum at around
$\lambda\sim 90.0$, while this should lowered to around $\lambda\sim 1.0$ by the $M_{IRS}$ 
preconditioner with convergence rate correspondingly improved by the reduced condition number of
the preconditioned operator.

\subsubsection{Alternative polynomial preconditioners}

Reusing the polynomial created by CG from a fixed number of iterations as a preconditioner has been previously 
studied \cite{Oleary}. In that study the problem arose that this polynomial, although Hermitian, may not be positive definite and damage the outer
convergence. The infra-red shift, as far as the author is aware first introduced in our work, greatly stabilises the CG polynomial 
and can be used to keep the polynomial sign definite over the range $[0,\infty)$ for even order polynomials. 
By keeping the preconditioner a data dependent Krylov process,
optimal under some norm, we appear to gain in stability.  Replaying the CG polynomial eliminates a single matrix multiply, and
also eliminates linear algebra in each iteration gaining around 10\% in runtime. On the other hand, 
it turns out that the dynamic response of CG to the spectral content of the residual
is also important  at the 10\%; running a fresh Krylov process for each preconditioner application (\emph{PreconditionerMirs})
results in faster convergence than freezing this with a static polynomial (\emph{PreconditionerMirsPoly}) and these are 
roughly competitive with each other.

We also compared to Chebyshev preconditioning. For the $M_{IRS}$ preconditioner the tuning problem is
vastly reduced since the upper end of the spectrum need not be identified. The Chebyshev preconditioner
is particularly sensitive to the high parameter since the polynomial explodes rapidly above
this threshold, while the optimal upper edge is found automatically by a CG based process. 
However, when well tuned the Chebyshev preconditioner can also almost as effective.

The two IRS preconditioners $M_{IRS}$ and $M_{IRSpoly}$ introduced in this paper are, as far as the author is aware, 
wholly new components of the algorithm.

\subsection{Robustness and flexibility}

We observe curious robustness effects during solution to $10^{-8}$ on a $16^3$ lattice in table~\ref{tab:dopdefl2}.
The preconditioned CG inverting the Schur complement operator $P_L {\cal H}$ is almost completely insensitive
to the precision of the preconditioner $M_{IRS}$ but is highly sensitive to errors in the little Dirac operator
inversion $Q=M_{SS}^{-1}$. Indeed convergence beyond the precision to which $Q$ is evaluated is not possible.

\begin{table}[hbt]
\begin{tabular}{c|c|c}
$M_{SS}^{-1}$ residual & $M_{IRS}$ residual & Iteration count \\
\hline
$10^{-11}$    & $10^{-8}$ & 36\\
$10^{-8}$    & $10^{-8}$  & Non converge
\footnote{smallest residual is $10^{-7}$ then diverges. Here
Luscher introduced flexible algorithms } \\ \hline
$10^{-11}$    & $10^{-8}$ & 36\\
$10^{-11}$    & $10^{-4}$ & 36\\
$10^{-11}$    & $10^{-2}$ & 36
\end{tabular} 
\caption{\label{tab:dopdefl2}
Sensitivity of the Schur complement inversion (DEF1) to both little Dirac operator residual,
and to the preconditioner residual. This corresponds to Luscher's algorithm, and it appears that
preconditioned CG is remarkably insensitive to preconditioner variability.
}
\end{table}

This confirms results in the numerical literature \cite{GolubYe}.
Although \emph{flexible} CG exists\cite{Notay} and could be used
to enhance tolerance to variability, we observe that CG is already 
surprisingly tolerant to variability in $M_{IRS}$ but not $Q$.
This may be understood as follows: in PCG,
the noise in the preconditioner $M_{IRS}$ \emph{only} enters the search direction, while
the linear combination coefficients entering the solution and residual update are based on 
matrix elements of $P_L {\cal H}$. 
In particular the multigrid papers also use the coarse grid operator as a preconditioner
and are less sensitive to convergence noise; it is this that admits the composition of many imprecise levels in a multi-grid
cycle scheme. One can certainly
conclude that it is better to use the little Dirac operator inverse as a preconditioner and not separate the solution into subspace and complement.

\subsection{Improved solver framework}
\label{sec:adef2}

To move the little Dirac operator into the preconditioner we extend framework of \cite{Tang} to three levels as follows. 
First we consider the most general possibilities for combining in a single preconditioner
a little Dirac operator $Q$ and $M_{IRS}$, each representing approximate inverses, where
$Q$ operates on a subspace containing almost all low modes and overlapping with a subset of high modes
(the splitting is necessarily \emph{inexact} due to the theta function that restricts
vectors to blocks), and $M_{IRS}$ is accurate on the high mode space.
Among the  options for combining these as a preconditioner in an outer solver is the naive additive case:
$
M_{IRS} + Q
$.
However, we can also consider alternating error reduction steps, such as:
\beq
\begin{array}{ccc}
x_{i+1} &=& x_i + M_{IRS} [ b - {\cal H} x_i]\\
x_{i+2} &=& x_{i+1} + Q [ b - {\cal H} x_{i+1}]\\
 &=& x_{i} + M_{IRS}[b-{\cal H} x_i] + Q [ b - {\cal H} [x_{i} + M_{IRS} [b-{\cal H} x_i] ]]\\
 &=& x_{i} + [(1-Q{\cal H}) M_{IRS} +Q ] (b-{\cal H} x_i) \\
 &=& x_{i} + [P_R M_{IRS} +Q ] (b-{\cal H} x_i)
\end{array}
\eeq
We can similarly infer the family of preconditioners listed in table~\ref{tab:precs} by choosing 
different sequences of error reduction steps.

\begin{table}[hbt]
\begin{tabular}{c|c|c}
Sequence & Preconditioner & Name\\
\hline
additive & $M_{IRS} +Q $ &AD\\
\hline
$M_{IRS}$, $Q$ & $P_R M_{IRS} +Q $ &A-DEF2\\
\hline
$Q$, $M_{IRS}$ & $ M_{IRS} P_L +Q $ &A-DEF1\\
\hline
$Q$, $M_{IRS}$, $Q$ & $P_R M_{IRS} P_L +Q $ &Balancing Neumann Neumann (BNN)\\
\hline
$M_{IRS}$, $Q$, $M_{IRS}$ & $M_{IRS} P_L + P_R M_{IRS} + Q - M_{IRS} P_L {\cal H}  M_{IRS}$ & MG Hermitian V(1,1) cycle
\end{tabular}
\caption{\label{tab:precs} In fact, a whole family of preconditioners arrived at by composing the little Dirac operator
and the IRS preconditioner. The IRS preconditioner can thought of as a
smoother in the multigrid context. For a HPD outer solver algorithm, multigrid requires
to use the V(1,1) cycle in the final row to preserve Hermiticity. }
\end{table}

We take $Q= \left( \begin{array}{cc}
0 & 0 \\
0 & M_{SS}^{-1}
\end{array}\right)$ and $M_{IRS}=({\cal H}+\lambda)^{-1}$  (or substitute $M_{IRSpoly}$ as appropriate),
and implemented the generalised preconditioned conjugate gradient algorithm in figure~\ref{fig:pcg_general}
and with the complete set of preconditioning options documented in table~\ref{tab:algos}.
Of these, we find that A-DEF2 is the most numerically efficient because the little Dirac
operator inverse enters only once, and in the preconditioning step so that the solution can be made very approximate
(see previous section).
The matrices in A-DEF1 and A-DEF2 are not manifestly Hermitian and further work is needed to show that one can
expect convergence of an outer conjugate gradient.  In fact it turns out that
DEF1(Luscher), DEF2, A-DEF1, A-DEF2, and BNN are \emph{equivalent} up to the convergence precision of the
little Dirac operator for appropriate start vectors. Thus the convergence rates only differ in thier
sensitivity to imprecision in inner Krylov inversions. We may see this as follows.

\begin{figure}
\begin{center}
{\small
\fbox{
\begin{minipage}{0.4\textwidth}
\begin{enumerate}
\item $x$ arbitrary
\item $x_0 = V_{\rm start}$ 
\item $r_0 = b - {\cal H} x_0$
\item $y_0 = M_1 r_0$ ; $p_0 = M_2 y_0$
\item for iteration $k$
\item \hspace{1cm}$w_k = M_3 {\cal H} p_k$
\item \hspace{1cm}$\alpha_k = (r_k,y_k)/(p_k,w_k)$
\item \hspace{1cm}$x_{k+1}= x_k + \alpha_k p_k$
\item \hspace{1cm}$r_{k+1}= r_k - \alpha_k w_k$
\item \hspace{1cm}$\mathbf{y_k = M_1 r_k}$
\item \hspace{1cm}$\mathbf{\beta_k = (r_{k+1},y_{k+1})/(r_k,y_k)}$
\item \hspace{1cm}$\mathbf{p_{k+1} = M_2 y_{k+1} + \beta_k p_k}$
\item end for
\item $x = V_{\rm end}$
\end{enumerate}
\end{minipage}
}
}
\end{center}
\caption{\label{fig:pcg_general}
Generalised preconditioned conjugate gradient algorithm of Tang et al\cite{Tang}.
The choice of matrices $M_1$, $M_2$, $M_3$ and vectors $V_{\rm start}$ and $V_{\rm end}$
interpolate between Hermitian $V(1,1)$ multigrid, Luscher's Schur complement
method, and introduce several new approaches. Of these this paper focuses on the A-DEF2
method which minimises little Dirac operator overhead while keeping all inner
Krylov solves in the preconditioner, and proves to be the most robust. We have
also implemented ``inexact preconditioned CG''\cite{GolubYe} and ``flexible CG'' \cite{Notay}
variants of this algorithm and these generalisations are simple to implement and as a result
are not documented in this paper to avoid repetitition.
}
\end{figure}

\begin{table}[hbt]
\begin{tabular}{c|c|c|c|c|c}
Method & $V_{\rm start}$ & $M_1$& $M_2$& $M_3$ & $V_{\rm end}$ \\
\hline
PREC   & $x$    &  $M_{IRS}$    &   $\ident$   &  $\ident$     &$x_{k+1}$\\
AD     & $x$    &  $M_{IRS}+Q$  &   $\ident$   &  $\ident$     &$x_{k+1}$\\
DEF1   & $x$    &  $M_{IRS}$  &   $\ident$   &  $P_L$     &$Qb + P_R x_{k+1}$\\
DEF2   & $Qb + P_R x$     &  $M_{IRS}$  &   $P_R$   &  $\ident$   &$x_{k+1}$\\
A-DEF1 & $x$  &  $M_{IRS} P_L + Q$  &   $P_R$   &  $\ident$   &$x_{k+1}$\\
A-DEF2 & $Qb + P_R x$  &  $P_R M_{IRS} + Q$  &   $\ident$  & $\ident$   &$x_{k+1}$\\
BNN    & $x$  &  $P_R M_{IRS} P_L + Q$  &   $\ident$  & $\ident$   &$x_{k+1}$\\
Multigrid & $x$ &  $M_{IRS} P_L + P_R M_{IRS} + Q - M_{IRS} P_L {\cal H}  M_{IRS}$ 
&   $\ident$  & $\ident$   &$x_{k+1}$
\end{tabular}
\caption{\label{tab:algos}
Spectrum of possible choices for the generalised algorithm.
DEF1 generates a Schur complement inverter similar to Luscher's original algorithm, while
mutilgrid corresponds to the standard symmetric V(1,1) cycle. The matrices in A-DEF1 and A-DEF2
are not manifestly Hermitian and further work is needed before once can
expect convergence of an outer conjugate gradient.  If we use $M_{IRSpoly}$ then PREC
is a polynomial preconditioned conjugate gradient with zero setup cost. We find the best
algorithm is A-DEF2.
}
\end{table}

\subsubsection{Hermiticity proof for A-DEF2}

The Hermiticity of $M_1$ is clear for BNN but not A-DEF2. However, we will reproduce\cite{Tang} a
proof that for $V_{\rm start} = Qb + P_R x$ A-DEF2 is identical to BNN.

We have from $Q {\cal H} = (1-P_R)$,
\beq
Q r_0 = Q [ {\cal H} V_{\rm start} -b ] =  (1-P_R) [Q_b + P_R x] - Qb = P_R Q_b = 0,\eeq
and
\beq
Q {\cal H} p_0 = (1-P_R) [ P_R M P_L + Q ] r_0 = 0.
\eeq
We obtain induction steps:
\beq
Q r_{j+1} = Q r_j - \alpha_j Q {\cal H} p_j = 0,
\eeq
\beq
Q {\cal H} p_{j+1} = (1-P_R) [ P_R M P_L + Q ] r_j + \beta_j Q {\cal H} p_{j} = 0.
\eeq
We can similarly show $P_L r_0 = 0$ and $P_L {\cal H} p_0 = {\cal H} p_0$  so that
\beq
P_L {\cal H} p_{j+1} = {\cal H} P_R [ P_R M P_L +Q] r_j + \beta_j p_j = {\cal H} p_{j+1},
\eeq
and
\beq
P_L r_{j+1}= P_L r_j - \alpha_j P_L {\cal H} p_j = r_j - \alpha_j {\cal H} p_j = r_{j+1}.
\eeq
The consequence is that BNN then retains $P_L r_j = r_j $ in exact evaluation, and
the BNN preconditioning ($P_R M P_L r_j$) and A-DEF2 preconditioning ($P_R M r_j$) must remain
equivalent up to convergence error of the inner Krylov steps.
In fact ref~\cite{Tang} shows that DEF1(Luscher), DEF2, A-DEF1, A-DEF2, BNN are \emph{all} 
equivalent up to convergence, but they differ hugely in the sensitivity to this convergence
precision.

It is interesting to note that since 
the equivalence of ADEF-2 iterates to those of a Hermitian preconditioner is inductive, it
is a proof that can only be obtained in a Krylov approach where the outer iteration steps are known.
In this sense the reduction of the number of smoothing steps from two in the case
of the Hermitian V(1,1) multigrid cycle to one in the case of ADEF-2 does not appear
to be a step one can take in the conventional multi-grid approach.

\subsubsection{Bfm solver algorithm support}

The BFM implementation supports many configurable solver options for the outer solver, documented in table~\ref{tab:finesolver},
for the outer level Krylov solver. The outer solver search direction update may be varied to
implement the standard, inexact preconditioned, and flexible variants of preconditioned conjugate gradients. 
The PcgType controls which of the two level algorithms introduced by\cite{Tang} are implemented.

This completes the discussion of the finest grid, and in the following section
we now consider the optimised implementation of the second level as an inexact Krylov process.

\begin{table}[hbt]
\begin{tabular}{c|c|c}
\multicolumn{3}{c}{ Preconditioner controls }\\
Parameter & Value & Meaning \\
\hline
PcgType & PcgPrec & Single grid -- only use $M_{IRS}$ preconditioning\\
        & PcgAD   & Additive preconditioning algorithm\\
        & PcgDef1 & Def1 preconditioning algorithm\\
        & PcgDef2 & Def2 preconditioning algorithm\\
        & PcgADef1 & Adapted Def2 preconditioning algorithm\\
        & PcgAdef2 & Adepted Def1 preconditioning algorithm\\
        & PcgBNN   & Balancinng Neumann-Neumann preconditioner \\
        & PcgMssDef & Use only the little Dirac operator as preconditioner (no smoother)\\
        & PcgAdef2f & Adapted Def2 using single precision preconditioner \\
        & PcgV11f & Hermitian V11 cycle multigrid using single precision preconditioner \\
\hline
\multicolumn{3}{c}{ Search direction controls }\\
\hline
Outer Solver   & Pcg & Preconditioned CG \\
               & iPcg & Inexact preconditioned CG \\
               & fPcg & Flexible preconditioned CG  \\
\hline
\end{tabular}
\caption{\label{tab:finesolver}
Options controlling behaviour of BFM HDCG outer solver. The recommended algorithm is PcgAdef2f combined with Flexible CG 
which makes use of single precision acceleration and has robust convergence.
}
\end{table}

\section{Little Dirac operator treatment}
\label{sec:ldop}

In this section we report on three techniques used to alleviate the cost
of the little Dirac operator, and on several implementation issues.

\subsection{Reducing coarse operator overhead}

In our approach we limit the stencil of the Little Dirac operator by requiring each block dimension be $\ge 4^4$.
Since for Mobius Fermions $M_{ee}^{-1}$ is entirely non-local in $s$-direction, we let the blocks stretch the full extent of the s-direction.
These constraints leave the little Dirac operator as a sparse in 4d with next-to-next-to-next-to-nearest coupling, but which takes
\emph{no more than one hop in each direction}. For such blockings the coarse matrix \emph{only} connects to 80 neighbours:
\begin{center}
 ($\pm \hat{x}$), ($\pm \hat{y}$), ($\pm \hat{z}$), ($\pm \hat{t}$)\\
( $\pm \hat{x} \pm \hat{y}$), ( $\pm \hat{x} \pm \hat{z}$),  ($\pm \hat{x} \pm \hat{t}$) ,  ($\pm \hat{y} \pm \hat{z}$), 
( $\pm \hat{y} \pm \hat{t}$), ( $\pm \hat{z} \pm \hat{t}$) \\
( $\pm \hat{x} \pm \hat{y} \pm \hat{z}$),
( $\pm \hat{x} \pm \hat{y} \pm \hat{t}$),
( $\pm \hat{x} \pm \hat{z} \pm \hat{t}$),
( $\pm \hat{y} \pm \hat{z} \pm \hat{t}$)\\
( $\pm \hat{x} \pm \hat{y} \pm \hat{z} \pm \hat{t}$)\\
\end{center}

The underlying cost is reduced to \emph{only} ten times more than non-Hermitian system, however,
reducing to 4d has potentially saved an $L_s$ factor. The saving is overestimated as we likely require more vectors to describe 5th 
dimension. In our implementation of the little Dirac operator for BlueGene/Q \cite{haring2012ibm,Boyle:2012iy}
good efficiency was required since the BAGEL/Bfm implementation\cite{Boyle:2009bagel}
of the fine operator performs around 10x that of compiled code. Fortunately, dense matrix-vector operations were amenable to 
optimisation using vector intrinsics for the IBM xlc compiler achieving over $60$ Gflop/s in single precision. 
The communication latency associated with the 80 small messages
was reduced by around fifty fold using the SPI communications layer, and 
offloading the eighty packets to DMA can be performed in under ten microseconds.

The little Dirac operator inherits Hermiticity and sparsity from the fine Hermitian matrix.
The inverse of the little Dirac operator is applied by Krylov methods.  Conjugate gradient,
deflated conjugate gradient, ADEF1, ADEF2, and multi-shift conjugate gradient methods were 
implemented.

\subsubsection{Calculation of little Dirac operator matrix elements}

The non-local stencil makes it somewhat harder to determine the matrix elements 
\beq
A^{ab}_{jk} = \langle \phi^a_j| {\cal H} | \phi^b_k\rangle
.\eeq
In the nearest neighbour coupled 
non-Hermitian and unpreconditioned case it is cheap to determine all non-zero matrix elements.
In our non-Hermitian case the matrix elements between each blocked vector and the 80 nearby blocks
may always be computed with only 81 matrix multiplies using a Fourier trick 
as follows.  Other schemes are possible, such as coloring schemes, but this trick simplifies programming
when the stencil size of the coarse operator (3) does not divide the global coarse lattice size.

We create 81 complex phases  $z^s_b$ for each block $b$ and indexed by $s$. These are 
\beq z^s_b = e^{i p^s_\mu x_\mu^b}\eeq where the block $b$ has block coordinates $x_\mu^b$ in the coarse grid.
These phases correspond to low lying Fourier modes $p^s_\mu = n^s_\mu \frac{\pi}{ L_{\mu}} $ on the coarse grid with
up to one unit of momentum in each direction.  Hence $s$ indexes the $81=3^4$ neighbours of zero in momentum space
and this has the same dimension as the coarse grid stencil. For each $s$ and $\mu$, $n^s_\mu \in \{ -1, 0, 1\}$.
For each subspace vector $\phi_k$ and Fourier mode $s$, we compute vectors containing these phases multiplying each sub-block
\beq|\tilde{\phi}_k(p^s)\rangle = \sum_b e^{i p^s_\mu x^b_\mu} | \phi^b_k\rangle.\eeq 
We apply the Hermitian matrix and construct matrix elements for each Fourier mode $p^s$ as follows
\beq e^{-i p^s_\mu x^b_\mu} \langle \phi_{k^\prime}^b |  {\cal H} | \tilde{\phi}_k(p^s)\rangle 
=
 \sum_{l\in {\rm stencil}} e^{i p^s_\mu \delta^l_\mu } \langle  \phi_k^b | {\cal H} |   \phi^{b+l}_k \rangle
=
\sum_{l\in {\rm stencil}} M_{sl} \langle  \phi_k^b | {\cal H} |   \phi^{b+l}_k \rangle,
\eeq 
where $\delta^l_\mu$ is the coordinate space translation, in coarse grid coordinates, associated with element
$l$ of the stencil. Having assembled the matrix elements of the 81 Fourier modes we can invert the matrix
\beq
M_{s l} =  e^{i p^s_\mu \delta^l_\mu }
\eeq
and form the matrix elements as 
\beq
\langle  \phi_k^b | {\cal H} |   \phi^{b+l}_k \rangle
= 
M^{-1}_{ls} e^{-i p^s_\mu x^b_\mu} \langle \phi_{k^\prime}^b |  {\cal H} | \tilde{\phi}_k(p^s)\rangle 
\eeq
This inversion can be performed sequentially in the four dimensions, similar to a multi-dimensional Fourier
transform.

\subsection{Little Dirac operator solver}

The BFM implementation supports many configurable solver options, documented
for the inner level Krylov solver in table~\ref{tab:ldopsolver}. 
\begin{table}[hbt]
\begin{tabular}{c|c|c}
Parameter & Value & Meaning \\
\hline
LittleDopSolver & LittleDopSolverMCR & Modified conjugate residual\\
                & LittleDopSolverCG  & Conjugate gradient\\
                & LittleDopSolverDeflCG & Deflated conjugate gradient\\
& LittleDopSolverADef2 & A-Def2 algorithm\\
& LittleDopSolverADef1 & A-Def1 algorithm\\
\hline
\end{tabular}
\caption{\label{tab:ldopsolver}
Options controlling the BFM HDCG inner solver algorithm. The recommended algorithm is 
LittleDopSolverAdef1 which 
uses the truncation of the little Dirac operator to nearest neighbour as a 
smoother in combination with a second level deflation space.
}
\end{table}

\subsubsection{Little Dirac operator deflation}

We then obtain a further speed up by deflating the deflation matrix, 
making the algorithm hierarchical.
We compute the second level of global vectors (with no further blocking) in the 
deflation hierarchy using either three 
steps of inverse iteration with a shifted matrix, 
or the 4th order rational filtering technique.
We produce around 128 deflation vectors in addition to the
original fine grid subspace vectors. 
We again find that multi-shift inversion is more cost effective.

These vectors are used to augment the set of easily obtained global deflation vectors 
discussed in appendix A.3 of \cite{Luscher:2007se}. 

Although three levels are involved, 
the coarsest level is a single dense matrix.  
We diagonalise this basis to make by multiplication the third level operator cheap.
This pattern of grid structures is identical to that used in Luscher's original
algorithm \cite{Luscher:2007se}, but we augment the second deflation subspace beyond those vectors that are free to obtain.

The parameters in the BFM implementation used to control the generation
Table~\ref{tab:CoarseSubspaceParams}.

\begin{table}[hbt]
\begin{tabular}{c|c}
Parameter & Meaning \\
\hline
\emph{LittleDopSolverResidualSubspace} & Residual to use during subspace generation\\
\emph{LittleDopSubspaceRational} & Whether to use rational filtering (true) or inverse iteration (false)
\end{tabular}
\caption{\label{tab:CoarseSubspaceParams}
Parameters to HDCG for controlling the generation of the coarse grid deflation space.
The same low pass filter threshold is used for the little Dirac operator as for the fine dirac operator.
}
\end{table}

When we combine this deflation speed up with a relaxed convergence criterion
discussed in section~\ref{sec:adef2} we find around a 100 fold reduction in little Dirac operator overhead on 
$48^3$ simulations at physical quark masses, table~\ref{tab:dopdefl1}.
\begin{table}[hbt]
\begin{tabular}{c|c|c}
Precision & Hierarchical deflation & iterations \\
\hline
$10^{-7}$          & N & 4478 \\
$10^{-7}$          & Y & 250\\
$10^{-2}$          & Y & 63
\end{tabular}
\caption{
\label{tab:dopdefl1}
Reduction in little Dirac operator overhead enabled by a second level of deflation and
by a reduction in required precision that comes from using the little Dirac operator in a preconditioner.
}
\end{table}

\subsubsection{Truncated little Dirac operator as preconditioner}

We can also use the improved solver framework to accelerate the inversion of the 
little Dirac operator $Q$ beyond simple deflated CG. In particular, we may introduce
a cheap approximate inverse matrix acting on high modes of the little Dirac operator to 
augment the global vector deflation as follows. 

\begin{table}[hbt]
\begin{tabular}{c|c|c}
Hops & Frobenius norm & number terms\\
\hline
0    & 627 & 1 \\
1    & 6.2 & 9 \\
2    & 0.08 & 33\\
3    & 0.0007 & 65\\
4    & 0.00003 & 81
\end{tabular}
\caption{\label{tab:frobnorm} We display the Frobenius norm of the $N_{\rm vec}\times N_{\rm vec}$ matrix 
$A_{jk}^{bb^\prime}$  connecting
two blocks $b$ and $b^\prime$ separated by a distance of taxicab norm measured as a number of hops.
The matrices clearly fall rapidly with distance, while the number of such matrices grows less rapidly.}
\end{table}

The cost associated with the next-to-next-to-next-to-nearest neighbour Little Dirac operator 
stencil may be alleviated considerably. The coefficients of the matrix fall rapidly with distance,
table~\ref{tab:frobnorm}, and this may be exploited by truncating the matrix to finite range and using this in a preconditioner. 
Table~\ref{tab:frobnorm} shows that omitting these terms is a small perturbation to the Little Dirac operator and we
produce a range truncated matrix $Q_{\rm trunc}$
which includes only the zero and one hop terms, but remains Hermitian. 

When used as the preconditioner in our second level of deflation
$Q_{\rm trunc}$ is appealing because the cost of the truncated nearest neighbour 
preconditioner is nine times less than the cost of the  the unmodified next-to-next-to-next-to-nearest neighbour
little Dirac operator $Q$.

However, perturbing the smallest eigenvalues of the matrix $Q$ in an uncontrolled way is not necessarily wise since 
it could make the truncated matrix less well conditioned or even sign indefinite.
This problem can however be avoided because by applying a modest infrared shift we preserve positivity of the eigenvalues,
and protect the condition number. 

We can therefore construct high mode preconditioners for the little Dirac operator, and use
these in combination the second level of low mode deflation previously
discussed. We do this by using the ADEF1 algorithm as the for solver the little Dirac operator. 
Both the subspace vectors for the little Dirac operator $\phi_k$ and the vectors $A\phi_k$ can be stored, 
and since no further blocking is performed the ADEF1 algorithm is preferable because by precomputing
$A\phi_k$ we can build the preconditioner without applying the untruncated little Dirac operator.
This saves an extra matrix multiplication 
by the untruncated little Dirac operator solver $Q$ in each iteration  compared to ADEF2.

The removal of fill-in terms has been used for some time in incomplete Cholesky and incomplete
LU factorisation preconditioners to prevent cost growth. We apply a similar idea here to the enlargened stencil generated in CGNE. 
The inclusion of an infrared shift to protect condition number and
ensure complementarity of the the preconditioner in a multi-level algorithm is also a new aspect.

The Bfm implementation supports using either a Chebyshev polynomial preconditioner of $Q_{\rm trunc}$ 
or a fixed order conjugate gradient for $M_{IRS}(Q_{\rm trunc})$. 

\section{Results}

\label{sec:results}
As a test system we study a single configuration of the RBC-UKQCD  $48^3\times 96 \times 24$ ensemble with
pion mass $M_\pi=140$MeV and inverse lattice spacing $a^{-1}=1.73$ GeV on 1024 node rack of BlueGene/Q. This 
represents a physical light quark simulation on a large volume and is therefore of particular interest for Lattice
QCD simulations. In RBC-UKQCD's current analysis $10^{-4}$ precision is used for inexact
propagators in an all-mode-averaging analysis
\cite{oai:arXiv.org:1212.5542,Shintani:2014vja}, and  $10^{-8}$ precision is used for
exact propagators.

\subsection{Optimised HDCG solver parameters}

The algorithm has many parameters which must unfortunately be optimised for each ensemble. The algorithm is therefore very much 
not a black box algorithm in the style of conjugate gradient. After this optimisation we settled on the algorithmic parameters
in table~\ref{tab:optimised}, and found very significant performance gains.

\begin{table}[hbt]
\begin{tabular}{c|c|c}
\multicolumn{3}{c}{Geometrical controls}\\
\hline
Parameter & Value & Meaning \\
\hline
NumberSubspace & 64  & Number of subspace vectors\\
Block & 4,4,4,6,4,24 & Block size \\
SubspaceSurfaceDepth & 24 & 5th dimension depth retained in subspace\\
\hline
\multicolumn{3}{c}{First pass subspace controls}\\
\hline
SubspaceRationalLs & 24 & $L_s$ for first pass \\
SubspaceRationalLo &   0.0003 & Fourth order low pass filter\\
SubspaceRationalMass &   0.00078 & Mass for first pass \\
SubspaceRationalResidual &   1.0e-5 & Residual\\
\hline
\multicolumn{3}{c}{Second pass subspace controls}\\
\hline
SubspaceRationalRefine & True & Improve subspace in a second pass\\
SubspaceRationalRefineLo& 0.001 & First order low pass filter \\
SubspaceRationalRefineResidual & 1.0e-3 & Residual for refinement step\\
\hline
\multicolumn{3}{c}{Little Dirac operator deflation controls}\\
\hline
LdopDeflVecs & 160 & Vectors used to deflate little Dirac operator\\
LittleDopSubspaceRational & True &  Use rational filter for subspace generation\\
LittleDopSubspaceRational & False &  Use inverse iteration for subspace generation\\
LittleDopSolverResidualSubspace &   1.0e-7 & Residual during subspace generation \\
LittleDopSolverResidualInner    &   0.04   & Residual during solver\\
LdopM1control & LdopM1Chebyshev &  Use a Chebyshev approx inverse preconditioner\\
LdopM1Lo      & 0.5 & Low bound of Chebyshev\\
LdopM1Hi      & 45  & High bound of Chebshev\\
LdopM1iter    & 16  & Order of Chebyshev\\
\hline 
\multicolumn{3}{c}{Outer solver controls}\\
\hline
OuterAlgorithm & PcgAdef2f & Outer solver algorithm\\
OuterAlgorithmFlexible  & 1 & Outer solver uses flexible CG\\
Preconditioner& Mirs      & Use infra-red shift CG preconditioner\\
PreconditionerKrylovIterMax & 7 & seven iterations\\
PreconditionerKrylovShift &  1.0& Shift eigenvalues by 1.0 \\
\hline
\end{tabular}
\caption{\label{tab:optimised}
Options controlling behaviour of BFM HDCG solver after optimisation for simulation on $48^3$ configurations
at the physical point on a $a^{-1}=1.73$GeV ensemble.
}
\end{table}

Table~\ref{tab:hierarchy} gives the
grid hierarchy used with $48^3\times 96 \times 24$ five dimensional Fermion solution
arrived at after careful optimisation of the algorithm parameters.
The best numerical performance is obtained by using the Adef2f algorithm 
for the outer level solver and the Adef1 algorithm for the little Dirac operator solver. 
The preconditioner on the fine grid is a fixed number
of CG iterations with the Hermitian matrix and an infrared shift: $M_{IRS}({\cal H},{\rm iter}=7,\lambda=1.0)$.

For the coarse grid preconditioner we use a Chebyshev polynomial of 
the truncated little Dirac operator matrix $Q_{\rm trunc}$. The Chebyshev
approximation of order $16$ to $\frac{1}{x}$ over the range $[0.5,45]$ is applied to the
matrix $Q_{\rm trunc}$.
Since the cost of $Q_{\rm trunc}$ is around $\frac{1}{10}$th that of the untruncated matrix
$Q$, it is not surprising that optimisation favoured using a higher degree polynomial.

\begin{table}[hbt]
\begin{tabular}{c|c|c|c|c}
level & grid & block & Inversion algorithm & preconditioner\\
\hline
fine  & $48^3\times96\times 24$    & $4\times4\times6\times 4\times 24$ & ADEF2 & $M_{IRS}({\cal H},{\rm iter}=7,\lambda=1.0)$\\
\hline
coarse& $12\times12\times8\times24$  & - & ADEF1 & ${\rm Cheby}(Q_{\rm trunc},{\rm iter}=16,\lambda\in[0.5,45])$\\
\hline
global& $1^4$ & - & Dense prediagonalised &\\
\end{tabular}
\caption{\label{tab:hierarchy}
Hierarchy of grids selected for deflation of $48^3\times 96$ Mobius Fermion inversions at the physical point on a
1.75 GeV lattice. As with Luscher's original algorithm \cite{Luscher:2007se} we use a single coarser grid and global vector
deflation. However in our algorithm additional near null-space vectors are added to augment those trivially from the fine grid subspace
vectors. This is particularly helpful because the non-local nature of the Hermitian matrix is reflected in the little Dirac
operator cost.
}
\end{table}

\subsection{Performance}

We display the wall clock timings comparing HDCG performance to standard conjugate gradients (double precision and
restarted mixed precision) and to EigCG in table~\ref{tab:wallclock}.

\begin{figure}[hbt]
\includegraphics[width=0.5\textwidth]{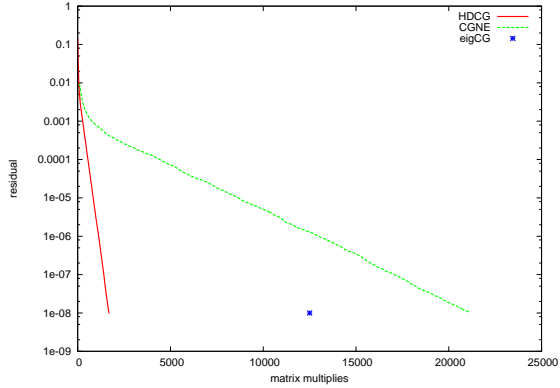}
\caption{
\label{fig:matmuls}
Residual versus fine lattice matrix multiples for the inversion of a gauge fixed wall source 
at physical light quark masses ($am=7.8\times 10^{-4}$) on a $48^3\times 96$ RBC-UKQCD
configuration with lattice spacing around $a^{-1}\sim 1.73 $GeV. We compare double precision conjugate gradient
on the normal equations to eigCG and to HDCG. HDCG is reduces the number of fine lattice matrix multiplies
by a factor of thirteen. 
}
\end{figure}

In figure~\ref{fig:matmuls} we show the convergence history of the HDCG algorithm on for the inversion of a gauge fixed wall source 
at physical light quark masses ($am=7.8\times 10^{-4}$) on a $48^3\times 96$ RBC-UKQCD
configuration with lattice spacing around $a^{-1}\sim 1.73 $GeV.
HDCG converged in 169 outer iterations and each outer iteration in HDCG used one double precision multiply and 
nine single precision multiplies.  Eight of these single precision multiplies were performed in the $M_{IRS}$ preconditioner.
Note that this comparison does not include two important effects: the single precision multiplies are significantly faster
to execute than double precision since the volume of data is halved, and the HDCG algorithm involves significant overhead
from an approximate inversion of the little Dirac operator in each iteration.

\begin{figure}[hbt]
\includegraphics[width=0.5\textwidth]{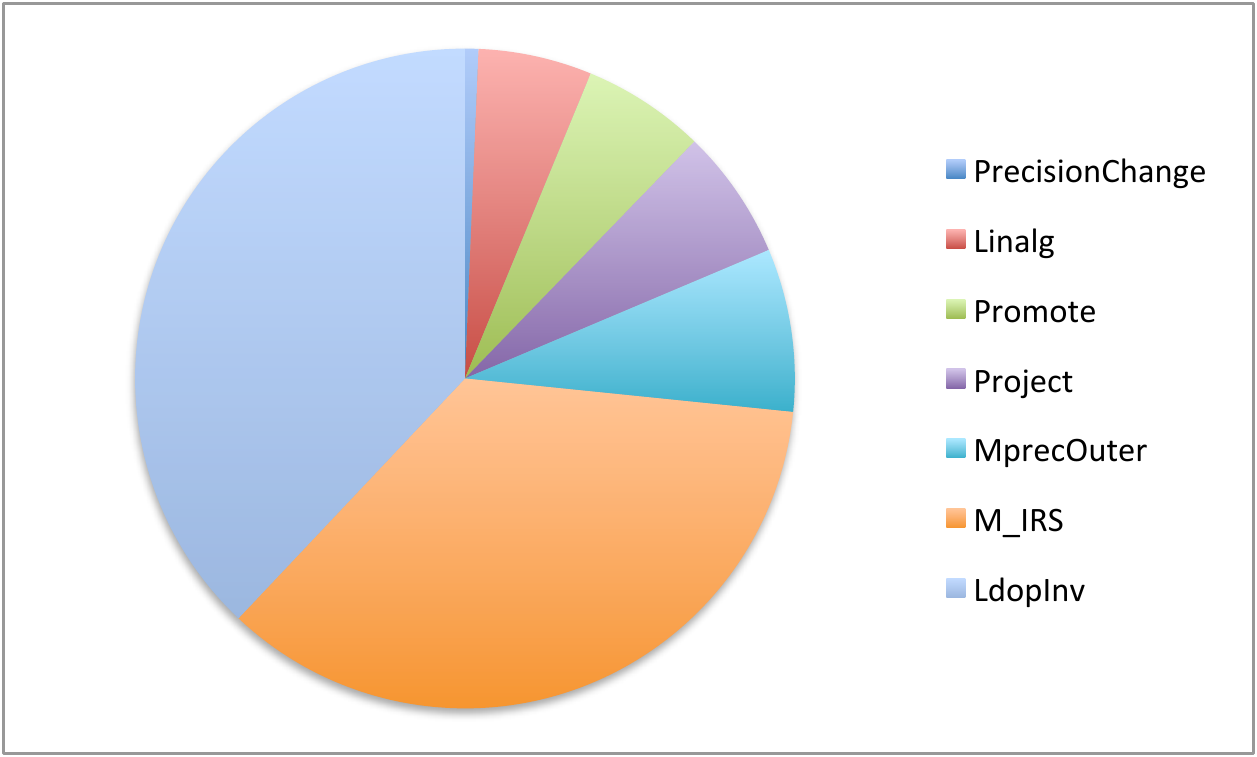}
\caption{
\label{fig:pie}
Components of the execution time of HDCG with the optimised parameters. Here, roughly half the time
is spent in applying the fine level matrix. This time comes from both single 
precision applications in the $M_{IRS}$ preconditioner
and double precision applications in the outer solver (labeled MprecOuter). 
Around one third of the time is spent solving the
little Dirac operator (LdopInv) and the rest of the time is made up from 
linear combinations, and the promotion and projection to and from the coarsened degrees
of freedom. Precision conversions constitute a small remnant contribution.
}
\end{figure}

\begin{figure}[hbt]
\includegraphics[width=0.5\textwidth]{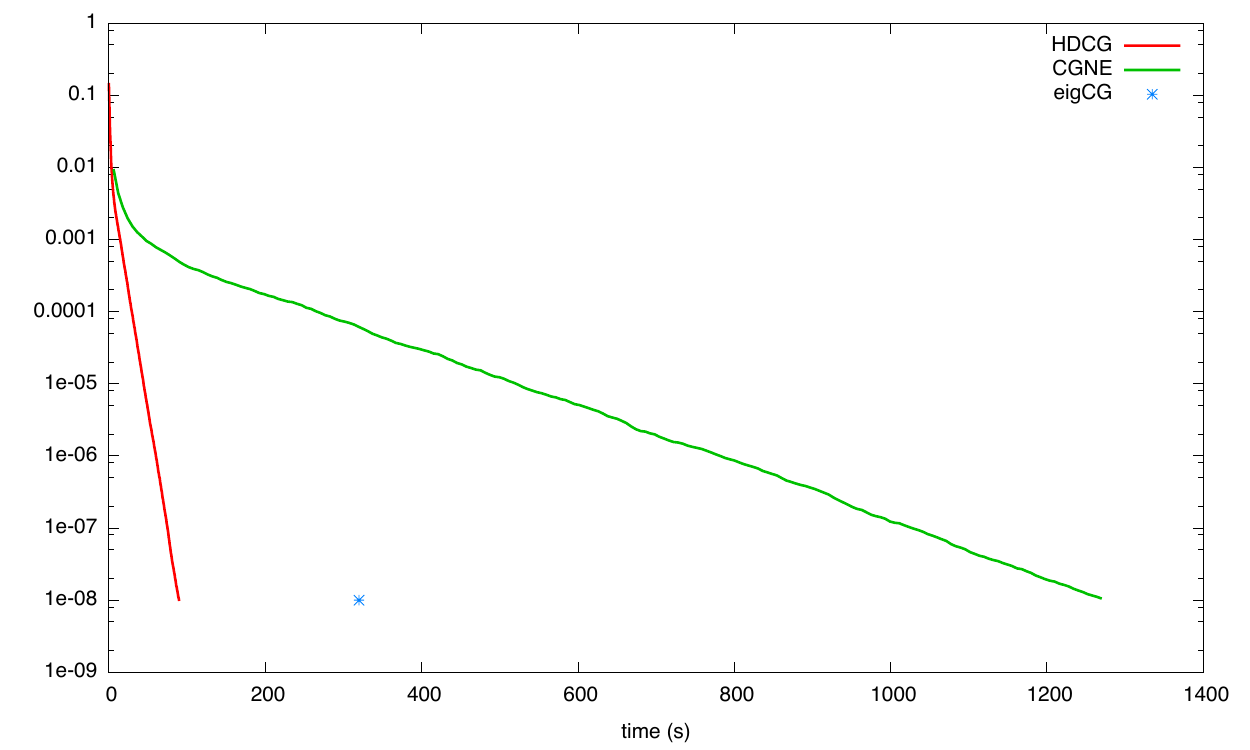}
\caption{
\label{fig:wallclock}
Wall clock execution time for the inversion of a gauge fixed wall source 
at physical light quark masses ($am=7.8\times 10^{-4}$) on a $48^3\times 96$ RBC-UKQCD
configuration with lattice spacing around $a^{-1}\sim 1.73 $GeV. We compare double precision conjugate gradient
on the normal equations to eigCG and to HDCG. A substantial speed up is obtained with HDCG algorithm
}
\end{figure}

The breakdown of the HDCG solve time is displayed as a pie chart in figure~\ref{fig:pie}.
As can be seen only around one half of the time is spent in fine matrix operations.
Including the extra overhead, we can compare the total wall clock execution time of the
different algorithms in figure~\ref{fig:wallclock}.  

Since the coarse space is a purely four
dimensional treatment and does not grow with $L_s$, it is clear that for sufficiently large
$L_s$ we can remove the preconditioner $M_{IRS}$ and achieve a cost effective solver that
does not grow with $L_s$. At least with the present algorithm
we are not at sufficiently large $L_s$ that this is a benefit, however the observation is
interesting.

In table~\ref{tab:speedup} we see that
with only 64 vectors we see a 3.5x speed up over EigCG, and a 14.1x speedup over double precision conjugate gradient applied
to the red black preconditioned normal equations. 
The setup time is substantially reduced compared to EigCG,
and the memory footprint is an order of magnitude reduced. The case for using the algorithm is rather compelling.

\begin{table}[hbt]
\begin{tabular}{c|cccc}
Algorithm & Tolerance & Cost & Matmuls & Vectors\\
\hline
CGNE (double)    & $10^{-8}$ & 1270s & 21144 & -\\
CGNE (mixed)     &  &  & 26000  & -\\
\hline
EigCG (mixed)    &  $10^{-8}$   & 320s & 11710 & 600\\
EigCG (mixed)    & $10^{-4}$    & 55s  & 1400  & 600\\
EigCG (setup)    &    & 10h   \\
\hline
HDCG (mixed)    & $ 10^{-8}$  & 90s & 1600 & 64\\
HDCG (mixed)    & $ 10^{-4} $ & 30s &  540 & 64\\
HDCG (setup)    &    & 2h &  \\
\end{tabular}
\caption{
\label{tab:wallclock}
We compare the performance of HDCG to existing algorithms for Mobius
and domain wall Fermion solvers. A substantial speed up is obtained compared
to both CGNE and to the deflated EigCG algorithm.
}
\end{table}

\begin{table}[hbt]
\begin{tabular}{c|c}
Comparison   & Gain \\
\hline
Exact Solve vs CGNE       & 14.1x \\
Exact Solve vs EigCG       & 3.5x \\
Inexact Solve vs EigCG       & 1.8x \\
Setup vs EigCG           & 5x \\
Footprint vs EigCG       & 10x 
\end{tabular}
\caption{
\label{tab:speedup}
We compare the performance of HDCG to existing algorithms for Mobius
and domain wall Fermion solvers. A substantial speed up is obtained compared
to both CGNE and to the deflated EigCG algorithm. 
}
\end{table}

\subsection{Convergence rate analysis}

We may estimate the spectrum of the Hermitian operator using the $M_{IRS}$ polynomial to estimate the
high end of the spectral range, and lowest modes of the diagonalised little Dirac operator to estimate
the low end.  

The lowest little Dirac operator eigenvalues treated in the second level delation lie in the 
interval $[2\times 10^{-5}, 2\times10^{-3}]$. 
The highest eigenvalues of the fine operator are of $O(100)$. We would therefore estimate a condition number of order $\kappa=5\times 10^6$.
The convergence factor bound corresponding to this condition number is
%
%
%
%
\beq
\sigma^{\rm predicted}(CGNE) = 0.9991.
\eeq
We measure the convergence rate from the asymptotic behaviour of CG by fitting fig~\ref{fig:matmuls} as
\beq
\sigma^{\rm measured}(CGNE) = 0.9994,
\eeq
corresponding to a condition number of $\kappa = 1.2\times 10^{7}$. Given the indirect connection between the
eigenvalues of our little Dirac operator and the true lowest mode of the fine Dirac operator this
level of consistency is quite reasonable.

We now make make the assumption that CG bound is saturated, and use this as a reasonably accurate guide
to the spectral radius of the deflated operator.
After composite preconditioning with both $M_{IRS}$ polynomial preconditioner and the little
Dirac operator we fit a convergence factor from fig~\ref{fig:matmuls}  as 
\beq
\sigma^{\rm measured}(HDCG) \sim 0.92
\eeq
This suggests a condition number of $\kappa \sim 600$. Since the $M_{IRS}$ preconditioner is
taken with $\lambda=1.0$, this the upper eigenvalue of the preconditioner matrix should lie
around unity. The condition number of the preconditioner matrix suggests that eigenvalues
down to around $2\times 10^{-3}$ are well treated by the little Dirac operator. This is again
quite consistent with the upper end of the spectral values observed in the deflation of the little Dirac operator.

In summary: we use two complementary preconditioners designed to accurately treat the low and high  ends of the spectrum
respectively. The little Dirac operator appears to lift the lowest eigenvector of the preconditioned system
by around two orders of magnitude from $2\times 10^{-5}$ to $2\times 10^{-3}$. The $M_{IRS}$ preconditioner
reduces the highest eigenvector of the preconditioned system by a similar amount from around $100$ to around
$1.0$. The combined effect is a two order of magnitude improvement in condition number and a reduction in 
outer iterations to around $160$. Each of these outer iterations, of course, involves around 10 matrix multiplies
to implement $M_{IRS}$, with most of these matrix multiples are performed in single precision. We will demonstrate
that even further reduced precision remains effective in this preconditioning in section~\ref{sec:sloppy}.

\subsection{Subspace reuse}

Since the subspace generation is a real investment of effort, it is interesting to consider situations where this 
investment can be reused. When the Dirac matrix is only modified by a small change, we have the opportunity to reuse the
blocked subspace vectors. The matrix elements of the modified Dirac operator may be recomputed and the deflation space reused.
Several types of modification might be considered as small perturbations for which this approach might be successful.
\begin{enumerate}
\item {\bf Twisted boundary conditions.} The Bloch theorem suggests that to leading order in the twist the eigenvectors
will simply acquire a slowly moving phase. By blocking the original vectors the large phase factors accumulated with
a large translation should be absorbed.
\item {\bf Moderate changes in Fermion mass.} This not obvious for five dimensional chiral Fermions as the mass is not
an additive shift of the spectrum.
\item {\bf Gauge fields modified by small timesteps} in a Hamiltonian update. This case has been considered by Luscher \cite{LuscherHMC} and found
effective. We note that this usage implies violation of reversibility at the level of convergence precision, rather than solely through rounding
error accumulation at the level of the floating point precision. However, since we already use single precision arithmetic in the molecular dynamics
phase of RHMC, and advocated further reduction of precision in a preconditioner (section \ref{sec:sloppy}) it is quite likely that in future
reversibility violation will be controlled by convergence precision in any case.
\end{enumerate}

Table~\ref{tab:twist} shows that one can indeed introduce twist angles to the gauge fields
with twist up to of order one unit of Fourier momentum in each direction without significant
loss of convergence. A further example can be obtained
from RBC-UKQCD's recent physical point calculation of the $K_{l3}$ form factor.
The momentum required to be injected to the final state pion was included with twisted boundary conditions,
and 90s was required for \emph{both} the twisted and untwisted inversions with a gauge fixed wall source.
This demonstrates that subspace reuse is highly effective with twisting.
Table~\ref{tab:twist} also shows that one can indeed effectively reuse the subspace between different quark 
masses.

\begin{table}[hbt]
\begin{tabular}{c|c|c|c|c}
Algorithm & Volume & mass & Twist & Solve time\\
\hline
CGNE &$32^4$ & 0.01 & $\frac{\pi}{L}(0,0,0)$            & 30s\\
\hline
HDCG &$32^4$ & 0.01 & $\frac{\pi}{L}(0,0,0)$            & 6.9s\\
HDCG &$32^4$ & 0.01 & $\frac{\pi}{L}(0.2,0,0)$          & 6.9s\\
HDCG &$32^4$ & 0.01 & $\frac{\pi}{L}(0.5,0.5,0.0)$      & 9.2s\\
HDCG &$32^4$ & 0.01 & $\frac{\pi}{L}(0.5,0.5,0.5)$      & 9.8s\\
\hline
HDCG &$32^4$ & 0.1  & $\frac{\pi}{L}(0,0,0)$            & 3.6s\\
HDCG &$32^4$ & 0.01 & $\frac{\pi}{L}(0,0,0)$            & 6.9s\\
HDCG &$32^4$ & 0.005& $\frac{\pi}{L}(0,0,0)$            & 7.4s\\
HDCG &$32^4$ & 0.001& $\frac{\pi}{L}(0,0,0)$            & 7.8s\\
\hline
\end{tabular}
\caption{
\label{tab:twist}
We investigate reuse of a subspace created with $m=0.01$ and with no twist in the boundary conditions for
inversion of the related linear systems of equations with modest twisted boundary conditions or with modest modifications
in the quark mass. These results are very encouraging.
}
\end{table}

\subsection{Reduced precision in preconditioner communication}
\label{sec:sloppy}
Our $M_{IRS}$ preconditioner does have the locality benefit of SAP; the excellent communication performance in BlueGene/Q tolerates this.
However motivated largely for future machines with less favourable communication performance we have investigated truncation of
the floating point mantissa to only 6 bits, by truncating each single precision word to 16 bits.  

Since BlueGene/Q has only IEEE floating point SIMD operations
this requires moving the data through the cache between floating point and integer register
files and applying a mask, rotate, or combine step; compression to 8bit is not cost effective since one is reduced
to byte operations. However, the decision of how best to invest years of effort and to spend millions of dollars in future depends
on definitively answering this question. This reduces the bytes per word to only two.
On architectures such Xeon Phi which possess \emph{both} floating point and integer SIMD operations truncation to 8 bit
is feasible using a sequence of SIMD operations: maxabs, divide, and convert to 8bit signed integer instructions. 
In this way a 24 element four spinor could be stored as a 16bit half precision prefactor and 24 8 bit signed integers.
This can potentially save a factor of eight in communication over a double precision implementation \cite{BrowerClarkeUseful}. 

Table~\ref{tab:sloppy} shows that this reduction in communication bandwidth is acheived with no algorithmic penalty in iteration
count. As a consequence this appears to be an attractive competitive approach to domain decomposition; rather than
suppressing communication entirely, only the most numerically significant parts of the communication are preserved.

\begin{table}[hbt]
\begin{tabular}{ccc|c}
Precision of inner communication & Exponent & Mantissa & Outer iteration count \\
\hline
64 bit     & 11 bit &  52 bit &168 \\
32 bit     & 8  bit &  23 bit &168\\ 
16 bit     & 8  bit &  7 bit  &168
\end{tabular}
\caption{
\label{tab:sloppy}
We compare applying the $M_{IRS}$ preconditioner with different levels of numerical precision.
The top two rows compare application with a uniform 64 bit and 32 bit precision of all elements of data.
The final row retains 32 bit precision for all elements of data except for data communicated between nodes.
The communication buffers are truncated to retain only seven mantissa bits, and no detrimental algorithmic 
impact is seen. Compared to the 64 bit case a four fold reduction in communication bandwidth has been achieved, however
in architectures where conversion to 8bit integer can be performed in SIMD instructions we can expect successful compression
by a factor near eight without loss of algorithmic efficiency.
}
\end{table}

Of course, the cache locality benefit of domain decomposition is not preserved here. However, for 5d chiral Fermions,
we obtain $L_s$ cache reuse of gauge fields and $2 N_d$ reuse of Fermion fields in the Dirac operator and there
is already a high level of cache reuse in the matrix multiply.

The greater cache locality of domain decomposition on current machines would still improve
the overall Krylov solver performance somewhat compared to HDCG since it would allow 
the linear combinations to be performed at cache bandwidth (rather than memory bandwidth).
However the reduction of communication bandwidth achieved in this section is by far the larger effect
particularly as memory technology is advancing more rapidly than interconnect technology.
Consequently, it appears likely that reduced communication precision will give most of the performance benefit 
of domain decomposition without introducing no loss of numerical inefficiency in an inner/outer solver.

\section{Conclusions}

In this paper we have developed an inexact deflation method to accelerating the red-black preconditioned normal equations.
The matrices studied have a forty times larger stencil required compared to the nearest neighbour non-Hermitian stencil that is
used with Wilson and clover Fermions. We introduced required several substantial algorithmic refinements,
in order to give a real speed up in algorithm running time.

The use of the ADEF-2 algorithm allowed improved robustness to loose convergence of
the little Dirac operator, but with no formal change in convergence compared to the Schur complement
algorithm first implemented in inexact deflation\cite{Luscher:2007se}. This reduced the little Dirac operator
overhead by a factor of ten. The preconditioned conjugate gradient solver is remarkably tolerant
to preconditioner variability, and this organisation of the matrices almost eliminates the need for
flexible algorithms.

The generation and use of additional deflation space vectors in a
hierarchical multi-level deflation further reduced the cost of the coarse space by a factor between three
and ten. The grid pattern here continues to mirror Luscher's original algorithm with a $1^4$ third grid. 

We introduced an infra-red shift preconditioner based on a fixed number of CG iterations to replace
the Schwarz procedure used in both Luscher's approach and in the multigrid papers.
This preconditioner has been demonstrated

A further factor of three reduction in little Dirac operator overhead was obtained by using an
infra-red shift preconditioner based on the truncation of our little Dirac operator to nearest neighbour.
This reduced the bulk of coarse grid matrix multiplies to the same stencil as in the Wilson case.

Since the coarse space is represented as a purely four dimensional system, we have perhaps taken an
important step towards alleviating $L_s$ scaling of 5d Chiral Fermions. 

\section{Acknowledgements}

All simulations in this work were performed on the STFC DiRAC BlueGene/Q facility in Edinburgh.
This work was supported by grants ST/K005790/1, ST/K005804/1, ST/K000411/1, ST/H008845/1, 
STFC Grant ST/J000329/1 and the European Union ITN StrongNET (Agreement 238353). 
The author particularly wishes to thank Martin Luscher, 
Mike Clark, Richard Brower, Andreas Juettner, Marina Marinkovic
and my colleagues in RBC and in UKQCD for useful conversations in both audible and
electronic forms.

\end{document}

%% file: pab.tex
\usepackage{graphicx}
\usepackage{bm}
\usepackage{multirow}
\usepackage{graphicx}
\usepackage{epsfig}
\usepackage{psfrag}
\usepackage{soul}
\usepackage{color}
\usepackage{multirow}
\usepackage{mathptmx}
\usepackage{mathrsfs}
\usepackage{amsmath, amssymb}
\usepackage{slashed}

\newcommand{\beq}{\begin{equation}}
\newcommand{\eeq}{\end{equation}}
\newcommand{\beqa}{\begin{eqnarray}}
\newcommand{\eeqa}{\end{eqnarray}}
\newcommand{\ident}{1}

\newcommand{\FslashA}[1]{\!\not{\hbox{\kern-2pt ${#1}$}}}
\newcommand{\FslashB}[1]{\!\not{\hbox{\kern+1pt ${#1}$}}}

\allowdisplaybreaks

\usepackage{dcolumn}